# Low frequency Raman spectroscopy of few-atomic-layer thick hBN crystals

I. Stenger[1], L. Schué[1,2], M. Boukhicha[3], B. Berini[1], B. Plaçais[3], A. Loiseau[2] and J. Barjon[1]

[1] Groupe d'Etude de la Matière Condensée, Université de Versailles St Quentin, CNRS, Université Paris-Saclay, 45 avenue des Etats-Unis, 78035 Versailles cedex, France

[2] Laboratoire d'Étude des Microstructures (LEM), CNRS, ONERA, Université Paris-Saclay, 29 avenue de la Division Leclerc, 92322 Châtillon, France

[3] Laboratoire Pierre Aigrain, Ecole Normale Supérieure, PSL Research University, CNRS, Université Pierre et Marie Curie, Sorbonne Universités, Université Paris Diderot, Sorbonne Paris-Cité, 24 Rue Lhomond, 75231 Paris Cedex 05, France

**Abstract**

Hexagonal boron nitride (hBN) has recently gained a strong interest as a strategic component in engineering van der Waals heterostructures built with two dimensional crystals such as graphene. This work reports micro-Raman measurements on hBN flakes made of a few atomic layers, prepared by mechanical exfoliation. The temperature dependence of the Raman scattering in hBN is investigated first such as to define appropriate measurements conditions suitable for thin layers avoiding undesirable heating induced effects. We further focus on the low frequency Raman mode corresponding to the rigid shearing oscillation between adjacent layers, found to be equal to 52.5 cm$^{-1}$ in bulk hBN. For hBN sheets with thicknesses below typically 4 nm, the frequency of this mode presents discrete values, which are found to decrease down to 46.0(5) cm$^{-1}$ for a three-layer hBN, in good agreement with the linear-chain model. This makes Raman spectroscopy a relevant tool to quantitatively determine the number of layers in ultra thin hBN sheets, below 8L.

Hexagonal boron nitride (h-BN) is a 2D insulator with a wide bandgap (>6eV) and an atomic structure very similar to that of graphene where boron and nitrogen atoms alternate at the vertices of a planar hexagonal sp$^2$ network. This material has recently attracted wide attention appearing as an excellent candidate for 2D-layers encapsulation and an ideal insulating substrate of graphene in 2D electronic devices due to its unique physical properties such as atomic flatness, chemical inertness, absence of dangling bonds or low dielectric screening. Further, its high radiative efficiency at an energy close to 6 eV offers new optic emission capabilities such as plane-emission display-type devices in the far ultraviolet [1]. Moreover, as shown by recent cathodoluminescence experiments, emission properties of few-layer hBN films are found to depend strongly on their thickness in the range 1 – 20 atomic layers (L) [2]. Nevertheless a bottleneck problem is the precise thickness determination, which is quite challenging in the case of hBN thin films. The usual approaches using optical contrast microscopy or atomic force microscopy applied to hBN films deposited onto appropriate SiO$_2$/Si substrates, are known to be very tricky for hBN, resulting in large uncertainties, most often exceeding one atomic layer [2, 3, 4, 5, 6].

Recently, Tan and coworkers [7] have put forward a novel approach to quantify the number of atomic layers of thin graphene flakes by Raman spectroscopy in the Ultra Low Frequency (ULF) region. This approach is based on the thickness dependence of a vibrational interlayer shear mode (ISM), specific to layered crystals and associated to the relative motion of atoms from adjacent atomic planes. This mode arises at very low frequency and presents discrete frequency values depending on the number of layers in ultrathin graphene flakes. This technique has already been successfully extended to other 2D materials such as MoS$_2$ [8, 9, 10, 11] and Black Phosphorus [12]. For hBN, the bulk value of the interlayer shear mode has been reported at 52.5 cm$^{-1}$ [13]. It is predicted to shift down to 38 cm$^{-1}$ for a 2L hBN (no shear mode for a monolayer) [10]. However, because of the non-resonant character of Raman scattering in hBN, Raman signal in hBN is weak - in contrast to the cases of graphene and transition metal dichalcogenides (TMDC). This is due to the large hBN band gap compared to accessible excitation energies in Raman spectroscopy [14]. Compensating this weakness by using high laser powers is risky as it can induce a sample heating as identified in our previous work [6].

In the present work, we report a detailed study of the Raman interlayer shear mode in hBN crystals down to 3L at low excitation power. Particular care is taken to avoid laser heating. To this end the temperature dependence of the hBN Raman spectrum is addressed first. The frequency shift observed at low thicknesses is further analyzed within the linear chain model.

The 2D crystals studied in this work are exfoliated from three bulk sources: a single crystal grown at high-pressure high-temperature (HPHT) provided by the NIMS (NIMS sample) [15], and two commercially available powders from the Saint-Gobain company TrèsBN®-PUHP1108 (Saint-Gobain sample), and from the graphene HQ company (HQ sample). The hBN sheets were obtained by exfoliation following the standard cleavage procedure and deposited onto silicon substrates covered by a thin silica layer, its thickness being optimally chosen for hBN flakes detection by optical contrast microscopy [2,6]. h-BN films thickness was determined from their optical contrast (OC) using the cross-correlated calibration procedure with Atomic Force Microscopy measurements as detailed in [2,6]. This procedure insures a thickness uncertainty within ± 0.2L in the range of 1-20L range [6]. The Raman scattering spectra were recorded in backscattering geometry using a high-resolution Raman spectrometer setup (Labram HR800 from HORIBA Jobin-Yvon). The laser plasma lines are removed by using a narrow bandpass filter and the Rayleigh emission line is suppressed by using three notch filters with a spectral bandwidth of approximately 5-10 $cm^{-1}$, making measurements possible down to 7 $cm^{-1}$. An Argon ion laser operating at 514.5 nm is focused onto the sample through a 100× objective with a numerical aperture of 0.80 to form an illumination spot of about 1 µm in diameter. A 1800 lines/mm grating enables us to have each pixel of the charge-coupled detector covering 0.57 cm$^{-1}$. In these conditions, a spectral resolution 0.7 cm$^{-1}$ is estimated from the width of the Rayleigh peak. As explained below, the laser power was kept below 5 mW to avoid sample heating. Crossed polarizers were used to reduce the spectrally broad background coming from inelastic light scattering of free carriers in the Si substrate [16]. The temperature was monitored into a Linkam oven. An argon gas flux was kept flowing over the sample surface to remove the low-frequency Raman modes from air. The acquisition times for the shear mode spectra varied from a couple of minutes to several hours, depending on the hBN thickness.

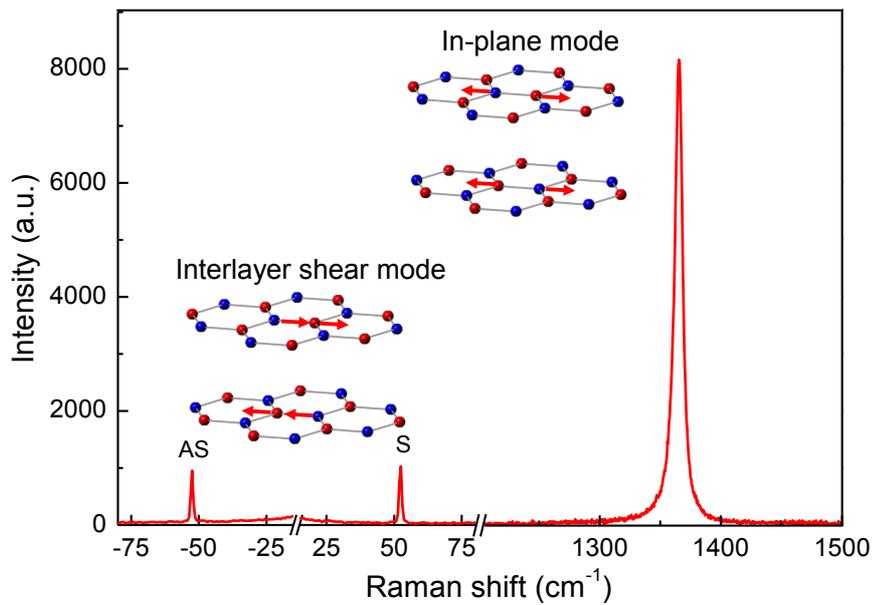

Fig. 1 .(one column width): Raman spectrum of bulk h-BN (NIMS sample).

We first consider the bulk situation, illustrated in Fig.1 by the typical Raman spectrum recorded on a NIMS crystal at room temperature. It exhibits the ultra-low frequency lines at -52.5 cm$^{-1}$ (anti-Stokes) and 52.5 cm$^{-1}$ (Stokes) with a full width at half maximum (FWHM) of about 1.1 cm$^{-1}$ and a high frequency line at 1366 cm$^{-1}$ (Stokes) with a FWHM of about 8 cm$^{-1}$. They respectively correspond to the interlayer shear mode (ISM) and the in-plane mode

(IPM), both having a $E_{2g}$ symmetry. No other line was observed in the 7 cm$^{-1}$ – 2000 cm$^{-1}$ range neither for the bulk flake nor for the thinner flakes. The integrated intensity of the 52.5 cm$^{-1}$ line is about the fiftieth of the 1366 cm$^{-1}$ line, making the low frequency mode much more difficult to detect. As both modes belong to the same irreducible representation, the large difference in intensity was interpreted as coming directly from the second derivatives of the polarizability tensor with respect to the normal coordinates (i. e. Raman tensor) [13]. It is the signature of the energy difference between the interlayer interaction - which is weak - and the in-plane atomic interactions - which are strong. Besides, contrary to the graphene and TMDC cases, Raman processes are non-resonant in h-BN when excitation laser sources are in the visible range. As a consequence, the ISM Raman signal from nanometer-thin hBN layers is much weaker than in other 2D materials. Recording it therefore requires longer integration times and special care to minimize the noise level. The use of a high laser excitation power is tempting, if not mandatory, but it can result in the sample heating inducing an additional frequency shift. We observed that heating effects are more pronounced in thin h-BN flakes on SiO$_2$/Si than in the bulk. Heating enhancement in low dimension structures was also reported in the case of hBN nanotubes [17].

We first address the effect of temperature on the Raman spectrum of bulk hBN in order to provide tools to rule-out heating effects in the analysis of atomically-thin hBN flakes. Both ISM and IPM modes were recorded on bulk hBN for temperatures ranging from 300K to 680K. The peak positions are fitted by using a Lorentzian function and their temperature dependences are reported in Fig. 2. As detailed in the supplementary information, the classical method to extract the sample temperature from the ratio of Stokes to anti-Stokes peaks is found to be not sensitive enough to detect and circumvent the heating problems. However, considering the temperature dependence of the Raman modes frequencies provides an advantageously efficient method. The data published in both hexagonal [18, 17, 19] and cubic [20] BN indicate a non-linear temperature dependence of the position of the IPM mode due to anharmonicity resulting from phonon–phonon interactions. In the temperature range of interest here, both ISM and IPM frequencies are found to vary linearly with temperature: slopes are evaluated to -0.006 cm$^{-1}$/K for the ISM and -0.023 cm$^{-1}$/K for the IPM respectively, consistent with those found in ref. [19]. A key point is that the frequency of the IPM is ~4 times more sensitive to the sample heating than the ISM one. It is then proposed to systematically follow the IPM mode together with the ISM to check the absence of undesirable heating. From this analysis, we have defined that faithful measurements conditions are secured when sample heating remains below 50K. In such conditions, frequency shifts of the ISM mode is within half of experimental resolution (0.3 cm$^{-1}$) whereas the shift of the IPM mode is clearly detectable but remains below 1.15 cm$^{-1}$. This approach is proposed as a perspective to reduce characterization times, particularly long in the range of a few atomic layers (several hours in this work at low laser power <5mW). We emphasize that the IPM frequency can be easily measured at low laser power down to 2L, thus producing the reference frequency for a given film. The IPM frequency is often reported to be slightly up-shifted by a few cm$^{-1}$ in few atomic layers films compared to the bulk material [3, 21], but showing no correlation with the number of layers. The absence of heating can thus be checked compared with the reference frequency taken at low laser power, no matter its value.

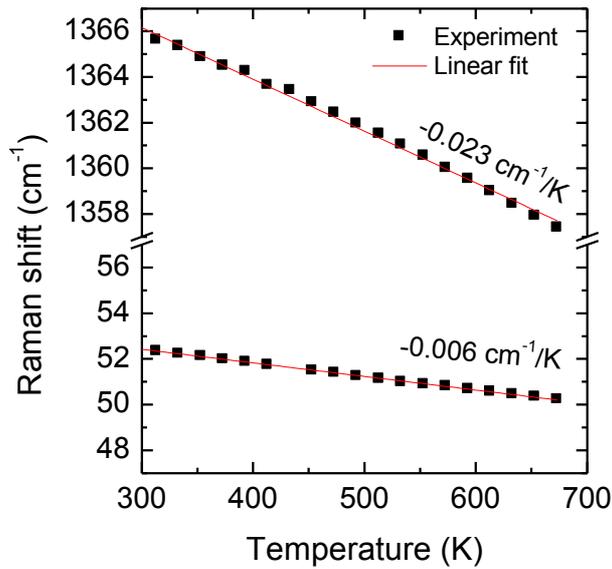

Fig2.(one column width, color online): Raman peak position of the ISM and the IPM as a function of temperature.

Having circumvented the heating problems, we focus on hBN flakes made of a few atomic layers. In Fig. 3a and 3b are plotted the Raman spectra for a set of samples of different thicknesses from 3L to 8L. The top spectrum is the reference bulk material. In Fig. 3a, we show that the IPM frequency does not change significantly as a function of the number of layers compared to the bulk, in good agreement with theoretical expectation [22,23]. Fig. 3b plots the antiStokes and Stokes parts of the spectra in the low frequency region. It shows that the position of the shear mode is strongly downshifted upon decreasing the hBN film thickness.

The dependence of the ISM frequency on the stacked layer number can be explained considering the simple linear-chain model proposed by Tan et al [7]. Let us consider a flake composed of N atomic hBN layers. The unit cell of a single hBN plane contains two atoms (boron and nitrogen). Assuming that the plane interacts only with its adjacent planes, the system can be reduced to a linear chain of N pairs of atoms bound along the c axis by N-1 springs as sketched in the insert of Fig3c. The shearing motion between planes can take place along two particular directions, armchair or zigzag, leading to the existence of two vibration modes. The frequency values expected for a hBN crystal made of N atomic planes can be determined from the relation :

$$\omega_\pm = \frac{1}{\sqrt{2}\pi c}\sqrt{\frac{\alpha}{\mu}}\sqrt{1 \pm \cos\left(\frac{\pi}{N}\right)}$$

with $\mu = 6.9\times10^{-27} kg.\text{Å}^{-2}$ the mass of a hBN plane per unit area and α the spring constant associated to the coupling between two adjacent planes. The value ω(∞) corresponds to the one measured in the bulk crystal with $\omega(\infty) = (1/\pi c)\sqrt{\alpha/\mu}$ = 52.5 cm$^{-1}$. We emphasize the absence of any adjustment parameter in the model. We deduce α = 16.9x10$^{18}$ N.m$^{-3}$. The values of ω$_+$(N) calculated for thickness varying from 2 to 20 monolayers are plotted in Fig3c. Obviously, only the highest frequency mode (ω$_+$) has been detected in our experiments. Secondary modes are also expected to appear [9,21], but none was detected for the moment.

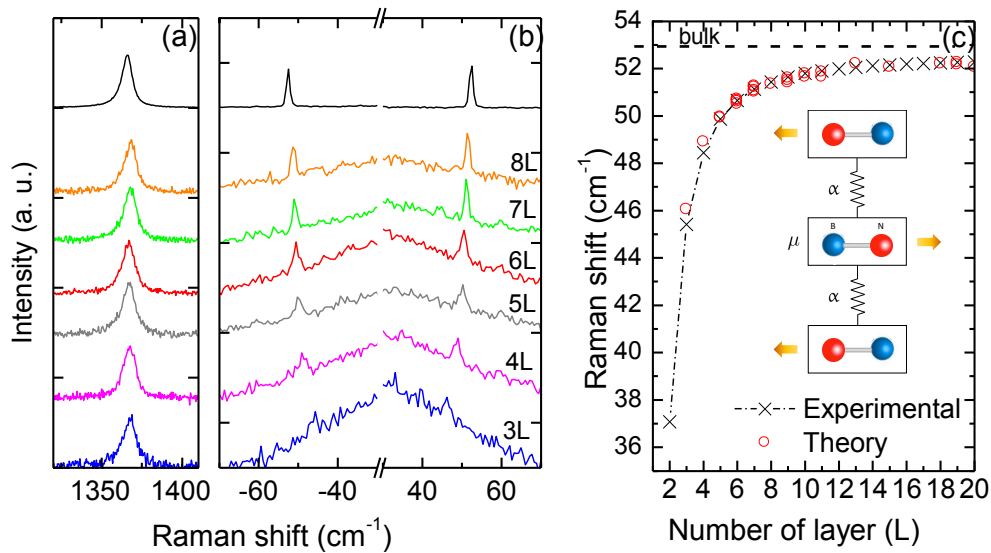

Fig. 3 (one column and a half width): Raman spectra of exfoliated hBN flakes having different thicknesses in the spectral region of: (a) the in-plane mode (IPM) (b) the shear mode (ISM). (c) Position of the ISM as a function of the atomic layer number: calculated with a linear chain model (cross), experimental (circle). The dashed line shows the bulk position at 52.5 cm$^{-1}$.

Experimentally, the ISM frequency is obtained as follows: Pos(ISM)=[Pos(ISM$_S$)-Pos(ISM$_{AS}$)] /2 where Pos(ISM$_S$) and Pos(ISM$_{AS}$) are the positions measured for the Stokes and anti-Stokes peaks respectively. The experimental positions obtained on 26 flakes are reported in Fig3c together with the linear-chain model. The agreement is excellent, thus definitely validating the model. For the thicker flakes, Pos(ISM) is shown to slightly decrease from 52.1 to 51.6 cm$^{-1}$ in the range 20 -9L. Below 8L, Pos(ISM) varies more strongly from 51.4(5) cm$^{-1}$ for a 8L flake down to 46.0(5) cm$^{-1}$ for a 3L flake. We further emphasize that some experimental points are superimposed, highlighting the discrete nature of the ISM frequency as the function of the number of hBN layers. To be specific, unique values are obtained for the flakes made of 6L (4 flakes), 7L (4 flakes), 9L (3 flakes), 2 flakes made of respectively 5L, 10L, 11L, 19L and 20L. For other number of layers, only one flake could be measured but with a specific value in each case. Although this sampling does not represent a large statistics, the repeatability of the measurements for the thinnest flakes is a good indication of the quantization of the ISM frequency. Taking into account our experimental resolution, the ISM frequency can thus be used to accurately determine the number of atomic layers in hBN films thinner than 8L.

In conclusion, we have used Raman spectroscopy to study the interlayer shear mode in few-layer hBN flakes. A quantitative recording procedure has been proposed to circumvent sample heating effects, particularly important in hBN. With these operating conditions, the in-plane mode frequency does not change consistently with thickness from the bulk down to 2L. On the contrary, we observe a large downshift and an accurate quantization of the shear mode frequency value for number of atomic layers below 8L. The linear chain model is validated and can be applied to determine the layer thickness of hBN flakes having a number of atomic layers down to 3 L.

The authors would like to thank the French National Agency for Research (ANR) for funding this work under the project GoBN (Graphene on Boron Nitride Technology), Grant No. ANR-14-CE08-0018. The research leading to these results has also received funding from the European Union Seventh Framework Program under grant agreement no. 696656 GrapheneCore1.

# Supplementary data

**Intensity ratio of the anti-Stokes on Stokes peaks vs temperature**

A classical indicator of sample heating is the intensity ratio between the anti-Stokes ($I_{AS}$) and Stokes ($I_S$) peaks which was measured in Fig. S1 for the IPM mode. It is in good agreement with theory where $I_{AS}/I_S = \exp(-\hbar\Omega/k_B T)$, with $\Omega$ the frequency of the vibrational mode, $k_B$ the Boltzmann constant [1]. In principle, this ratio could be used to evaluate the local temperature of a sample under laser exposure. However the sensitivity in the 300-400K range is weak as shown on. Fig. S1. The method exposed in the main body of the text was preferred.

Following the Stokes and anti-Stokes intensity of the ISM mode is not more reliable. In the low frequency region, the ratio $I_{AS}/I_S$ is close to unity and the variations with temperature are too low to be exploitable. Despite this approach has been suggested for graphene [1], we conclude that the ratio $I_{AS}/I_S$ of the ISM or IPM mode should not be used to evaluate the sample heating during Raman measurement.

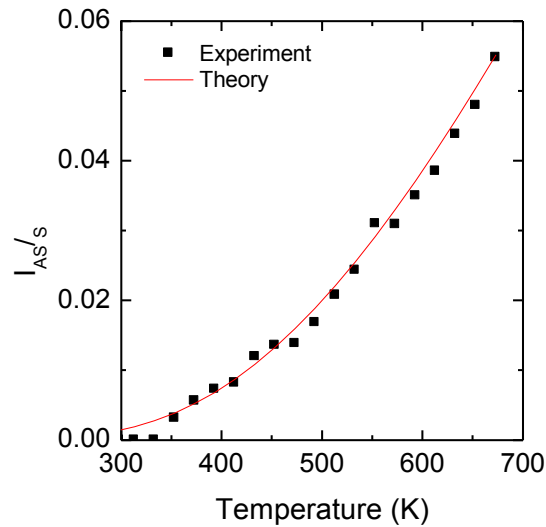

Figure S1: Intensity ratio between the anti-Stokes and Stokes peaks for the IPM mode at 1366 cm$^{-1}$ as a function of temperature.

## Thickness measurements

As pointed out in the main body of the text, the thickness determination is particularly difficult for hBN flakes in the atomic thickness range. An uncertainty below one atomic layer can only be achieved from the optical contrast (OC) when properly calibrated by Atomic Force Microscopy (AFM) on folded flakes measured on the same substrate. The procedure detailed in Ref. [2,3] insures a thickness uncertainty within ± 0.2L in the range of 1-20L. It was applied in this work, Fig. S2 showing optical images of atomically-thick layers measured with low-frequency Raman spectroscopy in the main body of the text (Fig. 3a and 3b). AFM on a folded flake used for OC calibration is shown in Fig. S3(a)(b) and the calibration relationship in Fig. S3(c).

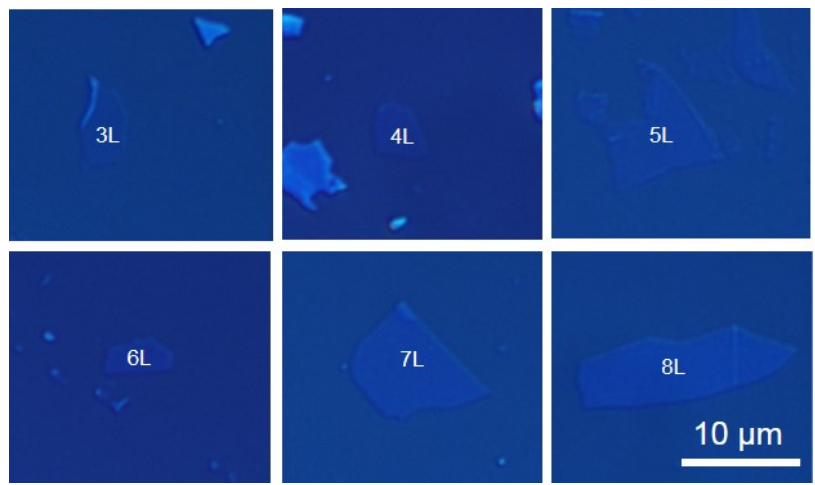

Figure S2: Optical images of hBN flakes measured by low-frequency Raman spectroscopy.

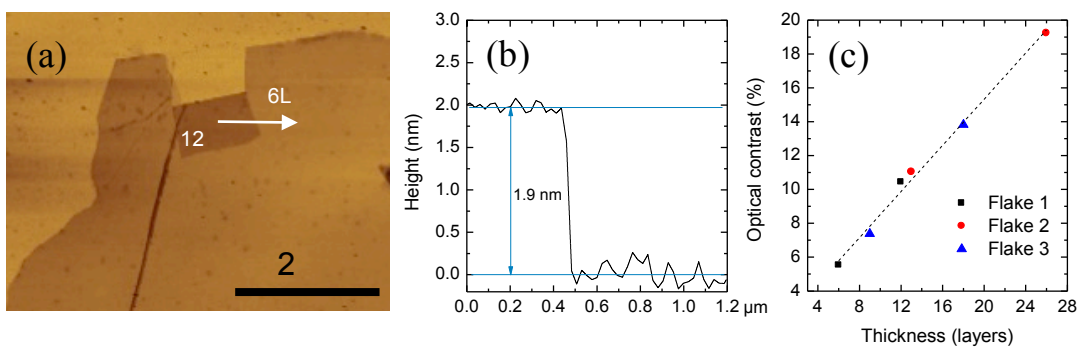

Figure S3 : (a) AFM image of a hBN folded flake used for OC calibration, (b) AFM profile taken along the arrow shown in (a), (c) Calibration plot of the optical contrast for three folded hBN flakes.